\documentclass[12pt, a4paper]{article}
\newtheorem{theorem}{Theorem}

\newtheorem{corollary}[theorem]{Corollary}

\begin{document}

\title{Game-theoretic discussion of \\
quantum state estimation and cloning}
\author{Chiu Fan Lee\thanks{c.lee1@physics.ox.ac.uk}
\ and \ Neil F. Johnson\thanks{n.johnson@physics.ox.ac.uk}
\\
\\ Center for Quantum Computation and Physics Department \\ Clarendon
Laboratory,
Oxford University \\ Parks Road, Oxford OX1 3PU, U.K.}

\maketitle

\abstract{
We present a game-theoretic perspective on the problems of quantum state
estimation and quantum cloning. This enables us to show why
the focus on universal machines and the different measures of success, as
employed
in previous works, are in fact legitimite.}

\newpage The topic of {\em quantum games} is a new area of study within
quantum
information, and its potential usefulness and consequences are still being
explored
and understood
\cite{Me99, EWL99, qgame,LJ02, LJ02t}. Early on in its development,
Meyer \cite{Me99} discussed a connection between quantum games and
quantum information processing. However the majority of the research carried
out on
quantum games to date, has lacked a direct connection to quantum information
processing.
 
In this paper, we discuss two quantum games which are of fundamental
interest to quantum information processing and specifically to
quantum crytopgraphy: the {\em quantum state estimation game}
and the {\em quantum cloning game}. These two quantum games are no strangers
to
physicists, and have been investigated extensively. However there are
still some significant questions waiting to be addressed. First: most work
in these
two areas focused on universal machines, and it was typically assumed that
the initial
pure state to be measured or cloned had a uniform distribution according to
the
corresponding unitary group. But why are these assumptions legitimate?
Second: the
measure of success in the cloning game was not unique, and researchers used
different
measures to facilitate their expositions. So why
did they end up with the same answers? The purpose of this paper is to
answer these
two questions by means of a unifying game-theoretic scenario which describes
these two
games.   

We start with a description of the two games concerned:

\noindent{\bf Quantum state estimation game}
Suppose there are two players, I and II, with a referee presiding over the
game. 
Player II chooses an arbitrary pure state
$|\psi\rangle
\in {\cal H}$ where
${\cal H}={\bf C}^d$. He then sends
$|\psi\rangle^{\otimes N}$ to player I and
$|\psi\rangle$ to the referee. Player I's task after receiving the $N$
qubits from player
II, is to perform a measurement on them. Based on the outcome, player I
sends a pure
state $|\phi\rangle \in {\cal H}$  to the referee.  For example, if player I
decides to
use the set of  POVM operators
$\{M_m\}$ to do the measurement, and if he associates  state
$|\phi_m\rangle$ to
measurement outcome $m$, the final state that he sends back to the referee
will be $\sum_m {\rm tr}[M_m \rho  M_m^\dag ]|\phi_m\rangle \langle \phi_m|$
where
$\rho =(|\psi\rangle \langle \psi|) ^{\otimes N}$. After receiving the two
qubits from
player I and player II, the referee applies the SWAP-test \cite{BCW01} on
them: if the test says that the two states are equal, he awards a payoff of
1 to
player I and $-1$ to player II. Otherwise, player I  gets a payoff of $-1$
and player
II a payoff of 1.

\noindent{\bf Quantum cloning game}
In this game, player II chooses a pure state
$|\psi\rangle
\in {\cal H}$ where
${\cal H}={\bf C}^d$. He sends the state
$(|\psi\rangle 
\langle \psi|)^{\otimes N}$ to player I
and the state
$(|\psi\rangle \langle \psi|)^{\otimes M}$
to the referee. After receiving the state from player II,
player I designs a device that takes as input $(|\psi\rangle \langle
\psi|)^{\otimes N}$ and outputs
a state $\sigma$ such that $\sigma$ is a density operator in ${\cal H}^{
\otimes M}$. He then sends $\sigma$ to the referee. Finally, the referee
applies the SWAP-test on $(|\psi\rangle \langle \psi|)^{\otimes M}$ and
$\sigma$.
If they pass
the test, he awards a payoff of 1 to player I and $-1$ to player II.
Otherwise he awards
$-1$ to player I and 1 to player II.

The above games are generalizations of the quantum games considered in
Refs.~\cite{EWL99, LJ02t}. However, one
fundamental difference from the games of Refs.~\cite{EWL99, LJ02t} is
that these games involve communication via qubits.
Moreover, comparing to the strategy sets discussed in Ref.~\cite{LJ02t},
we are placing a severe
restriction on the players' strategy sets. This is justifiable because only
one particular quantum operation is of interest.  We also note that
the quantum state estimation and quantum cloning
games that we consider here, are
equivalent to the games discussed in Ref.~\cite{MP95} and
Ref.~\cite{We98} respectively. The game-theoretic analysis which follows
below, however,
is new.

Before deducing any general
theorems for these games,  we will review some basic definitions in game
theory for
completeness. Further details are given in Ref.
\cite{osborne}. For a
vector
$\vec{v}=(v_i)_{i\in N}$ where
the $k$-th entry corresponds to the $k$-th player's
choice of strategy, we set
$\vec{v}_{-k}$ to be $(v_i)_{i\in N \setminus \{k \}
}$ and we denote $(v_1,
\ldots, v_{k-1}, v'_k, v_{k+1},
\ldots, v_N)$ by
$(\vec{v}_{-k}, v'_k)$.
We also define the set of so-called best replies
for player $k$ to be $B_k(\vec{\chi}_{-k}):=\{
\chi_k \in \Omega_k : P_k(\vec{\chi}_{-k},\chi_k) \geq
P_k(\vec{\chi}_{-k},\chi_k'), \forall \chi_k' \in \Omega_k \}$
where $P_k(\vec{\chi})$ is the payoff for the $k$-th player
given profile $\vec{\chi}$. We note that a strategy
$\chi_k \in B_k(\vec{\chi}_{-k})$
means that adopting $\chi_k$ renders the optimal
payoff to player $k$ if all the other players have chosen strategies
according to the strategy profile $\vec{\chi}_{-k}$.
Using this notion of best reply, we can easily define what a Nash
equilibrium is:
an operator profile $\vec{\chi}$ is a Nash equilibrium if
$\chi_k \in B_k(\vec{\chi}_{-k})$ for all $k$.

We are now ready to state and prove two theorems for these games:
\begin{theorem} 
\label{Minmax}
For the quantum state estimation and cloning games,
\begin{equation}
\label{P}
\max_{\chi \in \Omega_{\rm I}}
\min_{\xi \in \Omega_{\rm II}} P(\chi, \xi) =
\min_{\xi \in \Omega_{\rm II}}
\max_{\chi \in \Omega_{\rm I}} P(\chi, \xi)
\end{equation}
where $\Omega_{\rm I}$ and $\Omega_{\rm II}$ are the strategic sets for
player
{\rm I} and player {\rm II} respectively and $P(\chi, \xi)$ is the payoff
for player
{\rm I}.
\end{theorem}

\noindent {\bf Proof:}
We use the formalism developed in Ref.~\cite{LJ02t} where the same
theorem was proved. The only difference here is that we have allowed
an exchange of qubits between the two players, i.e.
the payoff matrix now has entries of the form:
\[
{\rm tr} [R (\underbrace{
\tilde{E} \otimes \tilde{E} \otimes I}_{{\rm I's \ operators}})
(\underbrace{
I \otimes \tilde{E} \otimes \tilde{E}}_{{\rm II's \ operators}})
\rho (\underbrace{
I \otimes \tilde{E}^\dag \otimes \tilde{E}^\dag}_{{\rm
II's \ operators}})
(\underbrace{
\tilde{E}^\dag \otimes \tilde{E}^\dag \otimes I}_{{\rm
 I's \ operators}})]
\]
with the summation indices omitted for clarity. We note that $R$ corresponds
to
the SWAP-test by the referee. Although the overall operation
on the initial state is no longer a direct product of the two players'
strategy sets, this detail is inessential with regards the theorem for the
following
reason: once a basis for the vector space of operators is fixed,
$\{\tilde{E} \}$
in this case, then all the coefficients attached to the operators
commute to the front. Hence one can see that the resulting payoff
function is again bilinear with respect to the vector spaces of the strategy
sets of the two players. Also, since we have allowed mixed strategies for
the
two players, the resulting strategy sets are convex. Compactness of the
strategy sets also follows from a similar discussion in Ref.~\cite{LJ02t}.
We 
can therefore deduce the theorem by invoking the
Minmax theorem of Ref.~\cite{Be97}
because the strategy sets are compact and convex, and $P(\chi, \xi)$
is linear and continuous in $\chi$ and $\xi$.
\hfill Q.E.D. \\ \\
\noindent The above proof is almost identical to the classical
version, but one should note that the two players' strategy spaces
no longer form a direct product of operator spaces.
\\

The next
theorem is more general than we need and it highlights the similarity
of the quantum state estimation and  quantum cloning games. In fact, the
same proof shows
that in the well-known children's game of rock-paper-scissors, the best
strategy
is to adopt the three options with uniform probability.

\begin{theorem}
\label{1}
Consider a two-player zero-sum game with pure strategy sets $\Omega_{\rm I}$
and $\Omega_{\rm II}$ for players I and II respectively,
such that the Minmax theorem applies.
Suppose that $\Omega_{\rm II}$ is a compact topological group with the
property that for each
$e \in \Omega_{\rm I}$, there is an $e_{f'} \in \Omega_{\rm I}$
such that the payoff $P(e_{f'},f)= P(e, f'f)$
for all $f \in \Omega_{\rm II}$.
Then the strategy $\xi^*$ which corresponds to adopting a specific strategy
with uniform probability with respect to the unique Haar measure, is
a strategy at a Nash equilbrium for player II.
\end{theorem}
{\bf Proof:}
We let $\chi^* \in B_{\rm I}(\xi^*)$ where $B_{\rm I}(\xi^*)$ is
the set of best replies of player I with respect to $\xi^*$.
$\chi^*$ can be represented by the probability ensemble $\{
Q(e)de,e \}$ where $Q(e)de$ is the probability for choosing the pure
strategy $e$.
We now define $\bar{\chi}^*$ to be the probability ensemble $\{
Q(e)\ de \ df',e_{f'} \}$ where $df'$ is a uniform probability
distribution
corresponding to the Haar measure of $\Omega_{\rm II}$. Then
\begin{eqnarray}
P(\bar{\chi}^*, \xi^*)&=&
\int de\ df\ df'\ Q(e) P(e_{f'},f) \\
&=&  \int de\ df\ df'\ Q(e) P(e,f'f) \\
&=&  \int de\ df\ Q(e) P(e,f) \\
&=& P(\chi^*, \xi^*).
\end{eqnarray}
Hence, $\bar{\chi}^*$ is also in $B_{\rm I}(\xi^*)$ but with the
extra property that $P(\bar{\chi}^*, f')= P(\bar{\chi}^*, f'')$
for all $f',f'' \in \Omega_{\rm II}$. Borrowing the terminology
from the cloning literature, we call $\bar{\chi}^*$ {\it universal}.
Now, suppose $\xi^*$ is not at a Nash equilibrium but $\hat{\xi}$ is
(its existence is guaranteed by the Minmax theorem). Then
\[
P(\bar{\chi}^*, \hat{\xi}) < P(\bar{\chi}^*, \xi^*).
\]
However this is impossible because $\hat{\xi}$ is itself a
probability ensemble of $\Omega_{\rm II}$. The theorem therefore follows.
\hfill Q.E.D. 
\\
\\
We can also deduce the following corollary from the above theorem:

\begin{corollary}
\label{cor}
Consider a game as depicted in Theorem~\ref{1}.
Any universal strategy $\bar{\chi}^*$, such that
$P(\bar{\chi}^*, \xi^*)$ is optimal and where
$ \xi^*$ is the same strategy as defined in Theorem~\ref{1},
is therefore at a Nash equilibrium.
\hfill {\rm Q.E.D.}
\end{corollary}

The quantum state estimation and quantum cloning games fall within the
realm of the previous theorems. We are therefore ready to discuss the two
questions
we set out to answer.
First: the assumption of having an initial pure state distributed
with uniform probability, is legitimate because this corresponds
to a strategy at Nash equilibrium for player II. Subsequently,
optimizing the estimation or cloning operation with respect to this strategy
gives
us the value of the corresponding game. Furthermore, if
the resulting operation is universal, then the operations are themselves
guaranteed to be at Nash equilibrium using the above corollary.
The reason why Werner obtained an identical result~\cite{We98} to
Gisin and Massar~\cite{GM97}, despite the fact that they
adopted different measures of success, is as follows: the quantity
Gisin and Massar identified for optimization (with $\chi \in \Omega_{\rm
I}$)
was
\begin{equation}
P(\chi, \xi^*),
\end{equation}
while that of Werner was
\begin{equation}
\inf_{\xi \in \Xi} P(\chi, \xi)
\end{equation} 
where $\Xi$ is the set of strategies whereby player II can only choose
one particular pure state
rather than a probabilistic mixture of many. Hence using
Theorem~\ref{Minmax} and Theorem~\ref{1}, we can see that
the bound that they arrived at is actually the
value of the cloning game. Furthermore, since Werner's optimization
was with respect to all unitary states, the operation found
is guaranteed to be at Nash equilibrium by Theorem~\ref{Minmax}.
On the other hand, since
the operation found by Gisin and Massar is universal, it is also
at Nash equilibrium due to Theorem~\ref{1}.

For completeness, we
now give the Nash equilibria for the quantum state estimation
and quantum cloning games:
\\
\\
The {\bf quantum state estimation game} with $N$ initial qubits
has a Nash equilibrium $(\chi^*, \xi^*)$
where 
$\chi^*$ corresponds to the following strategy:
\begin{enumerate}
\item
Measure the initial set of qubits with
the set of measurement operators
$\Big\{ \sum_{m,n} c_r e^{-i \psi_r(m-n)}
d^{N/2}_{m,N/2}(\theta_r) d^{N/2}_{n,N/2}(\theta_r)
|m\rangle \langle  n| \Big\}, 1 \leq r \leq N+1$ where the $c_r$'s are
such that
\begin{equation}
\sum_r c_r e^{-i \psi_r(m-n)}
d^{N/2}_{m,N/2}(\theta_r) d^{N/2}_{n,N/2}(\theta_r) = \delta_{m,n} \ .
\end{equation}
Here $|m \rangle $ is short-hand for $|\frac{N}{2}, m \rangle$ with
the principal axis adopted with uniform probability according to
the corresponding unitary group, and with
$d^{N/2}_{m,N/2}(\theta)$ being the
rotation operator of a spin-$N/2$ particle \cite{DBE98}.
\item
Upon measurement $s$, submit to the referee
the qubit $|\phi_s\rangle$ where
$|\phi_s \rangle^{\otimes N} = \sum_m e^{-i \psi_sm}
d^{N/2}_{m,N/2}(\theta_s) |m \rangle$.
\end{enumerate}
For player II, $\xi^*$ corresponds to adopting a pure state
with uniform probability with respect to the Haar measure of
the unitary group.
The value of the game is $\frac{N+1}{N+2}$.
\\
\\
The $N \mapsto M$
{\bf quantum cloning game} in a $d$-dimensional Hilbert space
has a Nash equilibrium $(\chi^*, \xi^*)$
where $\chi^*$ corresponds to the mapping \cite{We98}:
\begin{equation}
\label{cloning}
\rho \mapsto 
\frac{d[N]}{d[M]} s_M (\rho \otimes {\bf 1}^{\otimes (M-N)})s_M \ .
\end{equation}
Here $s_M$ is the orthogonal projection of ${\cal H}^M$
onto its Bose space
and
$\xi^*$ corresponds to adopting a pure state
with uniform probability with respect to the Haar measure of
the unitary group.
The value of the game is
$d[N]/d[M]$ where $d[N]=
\left( \begin{array}{c} d+N-1\\ N \end{array} \right)$.
\\
\\
\indent
Although
we have only succeeded in finding one particular strategy profile at Nash
equilibrium, this is in fact sufficient as far as playing the game is
concerned due to the following theorem:

\begin{theorem} In a two-player zero-sum game, let $(\chi_1, \xi_1)$ and
$(\chi_2,
\xi_2)$ be two equilibrium pairs. Then
\begin{enumerate}
\item
$(\chi_1, \xi_2)$ and $(\chi_2, \xi_1)$ are also equilibrium pairs, and
\item
$P(\chi_1, \xi_1) =P(\chi_2, \xi_2)=P(\chi_1, \xi_2)=P(\chi_2, \xi_1)$.
\end{enumerate}
\end{theorem}

\noindent {\bf Proof:} The proof can be found in
Ref.~\cite{Ow95}.
\hfill Q.E.D. \\
\\
In contrast to general games where one should worry about multiple Nash
equilibria, a strategy at equilibrium is as good as any other in a
two-player
zero-sum game. Therefore, finding one such equilibrium is enough.

We note that 
there is, in fact, more than one version of a quantum cloning game: these
versions differ
according to how the referee determines the payoffs. For example,
the referee may perform a one-particle test or multiple-particle test
on the qubits he/she receives. However it turns out that the strategies at
the
Nash equilibrium are equivalent \cite{We98, KW99}.
Unfortunately, their equivalence cannot be deduced from game-theoretic
arguments since the payoff vectors $R$ differ for these two games, and
consequently they are distinct in the game-theoretic formalism.
However, we can in fact deduce a bound on asymmetric cloning
from the Minmax theorem by considering the one-particle-test game.
In this game, the referee is going to apply
the SWAP-test on $|\psi \rangle \langle \psi |$, which is sent by
player II, together with a reduced density matrix $\sigma \in {\cal H}$
obtained
by tracing out all but the $k$-th Hilbert space ${\cal H}$ where
$k$ is chosen by player II.
In a similar way to the above argument,  one can see that Keyl and Werner
have shown that the
mapping in Eq.~\ref{cloning} is the strategy at Nash equilibrium, and the
value of a $N
\mapsto M$  cloning game is $\frac{N(d+M)+M-N}{(d+N)M}$.
Hence if there exists an asymmetric cloner such that
the sum of the fidelity of the $M$ output states with
respect to the original input state is greater than
$\frac{N(d+M)+M-N}{(d+N)}$,
then it will violate Theorem~\ref{Minmax}. Therefore {\it
given a $N \mapsto M$ cloner, the sum of the fidelity of the $M$ output
states with 
respect to the original input state is less than
or equal to $\frac{N(d+M)+M-N}{(d+N)}
=N +{\cal O}(1/d)$}. This bound thus limits the flow of information
from one system to another.
We note that the above bound on asymmetric cloning is also implicit
in Ref.~\cite{GM97}: however in that work the authors proceeded by
symmetrizing the asymmetric cloning machine.

In summary 
we have discussed the problems of quantum state estimation and cloning
using a game-theoretic perspective, and have found the corresponding Nash
equilibria. 
We also justified the focus to date on universal machines, and the different
measures of success employed. The fact that the theorems that we deduced are
more general
than we needed, implies that they have potential use in other
adversary-type scenarios. We also note that although
we have restricted the referee's action to be physical, hence rendering some
situations
impossible \cite{LJ02b}, this need not be the case.
In fact, the Minmax theorem holds as long as the
payoff function is rendered linear with respect to
$\chi$ and $\xi$. We conclude by noting that it is well-known among computer
scientists that bounds on classical computing can be proved by classical
game-theoretic
techniques \cite{MR95}. So could quantum games pay back this debt by passing
similar
benefits back over to quantum computation? The answer awaits
further investigation.

\vskip0.5in CFL thanks NSERC (Canada), ORS (U.K.) and the Clarendon Fund
(Oxford) for
financial support.

\end{document}